\documentclass[%
 reprint,
 superscriptaddress,
 amsmath,amssymb,
 floatfix,
 aps,
 prl
]{revtex4-2}

\usepackage{graphicx}
\usepackage{dcolumn}
\usepackage{bm}

\newcommand{\vect}[1]{\mathbf{#1}}
\newcommand{\matr}[1]{\mathbf{#1}}

\usepackage{mathtools}
\usepackage{mathrsfs}
\usepackage[colorlinks,urlcolor=blue,citecolor=blue,linkcolor=blue]{hyperref}
\usepackage{physics}

\begin{document}

\title{Phase Retrieval via Gain-Based Photonic XY-Hamiltonian Optimization}
\author{Richard Zhipeng Wang}
\author{Guangyao Li}
\affiliation{Department of Applied Mathematics and Theoretical Physics, University of Cambridge, Wilberforce Road, Cambridge CB3 0WA, United Kingdom}

\author{Silvia Gentilini}
\author{Davide Pierangeli}
\affiliation{Institute for Complex Systems, National Research Council (ISC-CNR), Via dei Taurini 19, 00185 Rome, Italy}

\author{Marcello Calvanese Strinati}
\affiliation{Research Center Enrico Fermi, Via Panisperna 89A, 00185 Rome, Italy}

\author{Claudio Conti}
\affiliation{Department of Physics, University Sapienza, Piazzale Aldo Moro 5, Rome 00185, Italy}

\author{Natalia G. Berloff}
\email[correspondence address: ]{N.G.Berloff@damtp.cam.ac.uk}
\affiliation{Department of Applied Mathematics and Theoretical Physics, University of Cambridge, Wilberforce Road, Cambridge CB3 0WA, United Kingdom}

\date{\today}
\begin{abstract}
    Phase-retrieval from coded diffraction patterns (CDP) is important to X-ray crystallography, diffraction tomography and astronomical imaging, yet remains a hard, non-convex inverse problem. We show that CDP recovery can be reformulated exactly as the minimisation of a continuous-variable XY Hamiltonian and solved by gain-based photonic networks. The coupled-mode equations we exploit
are the natural mean-field dynamics of exciton-polariton condensate lattices, coupled-laser arrays and driven photon Bose–Einstein condensates, while other hardware such as the spatial photonic Ising machine  can implement the same update rule through high-speed digital feedback, preserving full optical parallelism. Numerical experiments on images, two- and three-dimensional vortices and unstructured complex data demonstrate that the gain-based solver consistently outperforms the state-of-the-art Relaxed-Reflect-Reflect (RRR) algorithm in the medium-noise regime (signal-to-noise ratios 10–40 dB) and retains this advantage as problem size scales. Because the physical platform performs the continuous optimisation,  our approach promises fast, energy-efficient phase retrieval on readily available photonic hardware. uch as two- and three-dimensional vortices, and unstructured random data. Moreover, the solver's accuracy remains high as problem sizes increase, underscoring its scalability.
\end{abstract}
\maketitle

\section{Introduction}
\label{sec:intro}

Recently, there has been rising interest in using physics-inspired, physics-based computing systems for solving hard optimisation problems,  including many that are NP-hard~\cite{Nikita_AQT2023,lucas_ising_2014, syedPhysicsEnhancedBifurcationOptimisers2022b,honjo100000spinCoherentIsing2021, honari-latifpourCombinatorialOptimizationPhotonicsinspired2022, inagakiCoherentIsingMachine2016}. An important  example is the minimization of the \textit{XY} Hamiltonian with sign-varying couplings between spins, where each spin is allowed to take on a continuous value in the vector 
$
{\bf s}_i \;=(\cos \theta_i, \sin \theta_i)$ or complex form $s_i=e^{i \theta_i},$ where $\theta_i \in [0,\, 2\pi).
$
Such continuous-spin systems appear naturally in gain-dissipative photonic lattices, exciton-polariton networks, laser arrays, and related platforms, where each oscillator is represented by a complex variable whose amplitude and phase evolve in time~\cite{kalininPolaritonicNetworkParadigm2019, kalininNetworksNonequilibriumCondensates2018, partoRealizingSpinHamiltonians2020a, kimCombinatorialClusteringCoherent2024, pal2020rapid, nixon2013observing}.

The \textit{XY} minimization problem involves the quadratic Hamiltonian
\begin{equation}
    H_{\mathrm{XY}}(\{s_i\}) \;=\; -\tfrac{1}{2}\sum_{i,j}^N J_{ij}\, s_i\, s_j^{*}
    +\;\mathrm{c.c.},
    \label{eqn:xy_hamiltonian}
\end{equation}
where  \(J_{ij}\) represents  the pairwise couplings. The task is to find the spin configuration \(\{s_i\}\) that minimizes \(H_{\mathrm{XY}}\). This is a continuous quadratic optimization (QCO) problem, with applications ranging from clustering~\cite{kimCombinatorialClusteringCoherent2024} to portfolio optimization~\cite{wangEfficientComputationUsing2025}.

One key motivation for studying \textit{XY}-type physical networks is their ability to perform highly parallel, analogue searches for low-energy configurations, thereby offering an alternative to purely digital algorithms. Recently, there have also been efforts to force an \textit{XY}-based system into effectively binary (Ising-like) states by introducing a large penalty term. Specifically, one adds
\begin{equation}
    H_P(\{s_i\}) \;=\; H_{\mathrm{XY}}
    \;+\; P\sum_{i=1}^N\bigl[s_i^{*2} + \mathrm{c.c.}\bigr],
    \label{eqn:ising_hamiltonian_penalty}
\end{equation}
where \(P>0\) is chosen so as to penalize any non-zero real part of \(s_i\). If \(P\) is sufficiently large, spins tend to align at phases \(0\) or \(\pi\), effectively reducing the continuous-spin \textit{XY} problem to the binary Ising problem. 

The continuous nature of the \textit{XY} Hamiltonian also makes it  relevant to other QCO tasks. One such problem of practical importance is the phase retrieval problem. This is typically stated as follows: given a real measurement vector \(\mathbf{b} \in \mathbb{R}^M\) and a complex matrix \(\mathbf{A} \in \mathbb{C}^{M \times N}\), one seeks to find a complex vector \(\mathbf{x} \in \mathbb{C}^N\) satisfying
\begin{equation}
    \lvert \mathbf{A}\,\mathbf{x}\rvert \;=\;\mathbf{b},
    \label{eqn:phase_retrieval}
\end{equation}
where \(\lvert \cdot\rvert\) denotes element-wise amplitude.
The phase retrieval practical significance arises because this problem is often encountered in applications such as X-ray crystallography \cite{harrisonPhaseProblemCrystallography1993, barnettAnalysisCrystallographicPhase2024}, astronomical imaging \cite{kristPhaseretrievalAnalysisPre1995}, and diffraction imaging \cite{miaoExtendingMethodologyXray1999, shechtmanPhaseRetrievalApplication2015}. 

In many of these applications, the complex vector \(\mathbf{x}\) represents the complete information about the sample and is referred to as the \emph{sample vector}. The matrix \(\mathbf{A}\) describes the action of the optical system, often well-approximated by a Fourier transform. While the intensity of the resulting electromagnetic wave can be measured by standard detectors (e.g., charge-coupled devices) to yield the real-valued amplitude \(\mathbf{b}\), the phase component is typically lost in the measurement process. Recovering \(\mathbf{x}\) from \(\mathbf{b}\) and \(\mathbf{A}\) constitutes the \emph{phase retrieval problem}, which is NP-complete~\cite{huangNoExistenceLinear2025}, underscoring its computational difficulty.

Moreover, without further constraints, the phase retrieval problem is frequently \emph{ill-posed} because multiple distinct sample vectors \(\mathbf{x}\) can give rise to the same measured amplitude \(\mathbf{b}\). For instance, if \(\mathbf{A}\) is a square matrix representing a discrete Fourier transform, any arbitrary phase profile can be applied to \(\mathbf{b}\) before performing an inverse Fourier transform, producing infinitely many valid solutions \(\mathbf{x}\). In such scenarios, even a theoretically exact algorithm may yield a recovered vector \(\mathbf{\tilde{x}}\) that deviates from the original \(\mathbf{x}\)~\cite{eldarPhaseRetrievalStability2014}.
This means that any solution of the form $\vect{x} = \matr{D}^{(i)} \matr{A}^{-1} \vect{b}$, where $\matr{D}^{(i)}$ is a diagonal matrix with arbitrary diagonal elements of the form $e^{i\theta_i}$, $\theta_i \in [0, 2\pi)$ and $\matr{A}^{-1}$ is the inverse discrete Fourier transform matrix, will satisfy the requirement given by Eq.~\eqref{eqn:phase_retrieval}.
Efforts to mitigate ill-posedness often rely on additional assumptions about the sample vector \(\mathbf{x}\). For instance, some approaches impose a sparsity constraint by specifying the number of nonzero entries in \(\mathbf{x}\)~\cite{bendoryAutocorrelationAnalysisCryoEM2023}. Others adopt even stricter conditions, stipulating the exact support of \(\mathbf{x}\); that is, which elements are nonzero~\cite{fienupPhaseRetrievalAlgorithms1982, chenApplicationOptimizationTechnique2007, latychevskaiaIterativePhaseRetrieval2018}. However, none of these methods can guarantee that the resulting phase retrieval problem is well-posed (i.e., admits a unique solution).

\begin{figure}
    \centering
    \includegraphics[width=\columnwidth]{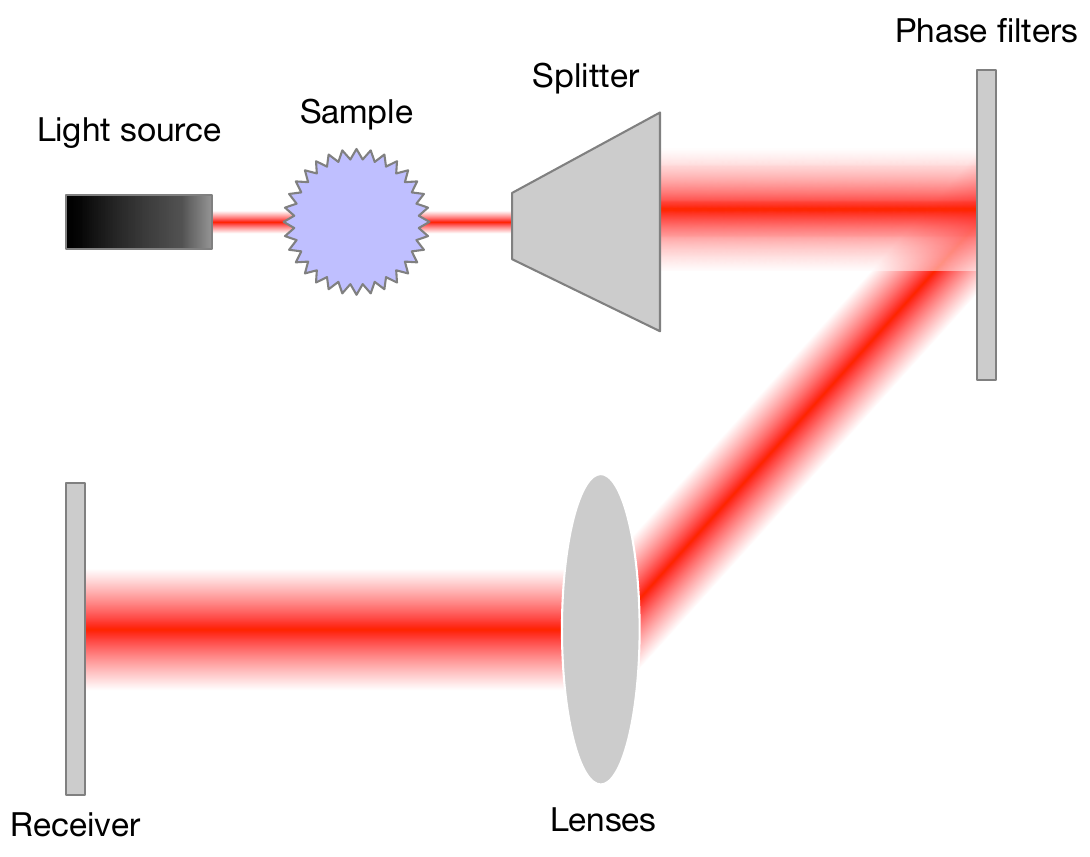}
    \caption{
    A schematic diagram for the CDP experiment framework.
    A light source is diffracted by the sample under investigation, and the diffracted complex-valued signal is split into $L$ identical beams, each of which is directed towards a phase filter that modifies the phase of the incident signal at each spatial location.
    The phase-modified beams are then directed through a lens system, and their intensities are finally captured.
    }
    \label{fig:0_cdp_schematic}
\end{figure}

Candes et al. \cite{candesPhaseRetrievalCoded2015} proposed a formulation called the ``Coded Diffraction Pattern'' (CDP) experiment, which provides some solution uniqueness guarantee.
Under this formulation, a sample vector $\vect{x}$ is first subjected to the action of $L$ random phase filters, each defined by a diagonal matrix $\matr{D}^{(i)}, i=1,...,L$. 
Each filter is defined by diagonal entries that are complex numbers with unit magnitude and randomly distributed phases. Thus, the filter modifies only the phase of the incoming signal while leaving its amplitude unchanged.
This produces $L$ filtered sample vectors, which are Fourier transformed separately.
The results are then concatenated, and amplitudes are taken to form the observation vector $\vect{b}$.
Hence, in this formulation, $\matr{A}$ is a $M$ by $N$ complex matrix, where $M=L \times N$, and represents the combined action of the random filters and the Fourier transforms.
Fig.~\ref{fig:0_cdp_schematic} provides a schematic diagram showing the experimental setup of this arrangement.
This formulation is theoretically appealing because  that if the distribution of the random diagonal matrix elements satisfies certain conditions, then with sufficiently large $L$, the solution of the phase retrieval problem is guaranteed to be unique Ref.~\cite{candesPhaseRetrievalCoded2015}.
It was also demonstrated numerically that even with filters whose diagonal matrix elements did not follow random distributions that satisfy the conditions laid out in \cite{candesPhaseRetrievalCoded2015}, the problem could still become well-posed with increased $L$.
Empirically, the process of applying multiple filters corresponds to the practice of oversampling the target and is reasonably achievable in many imaging applications \cite{candesPhaseRetrievalCoded2015, fannjiangPhaseRetrievalRandom2012, fannjiangNumericsPhaseRetrieval2020}.

The performance of conventional phase retrieval algorithms in this setting has been systematically examined in \cite{fannjiangPhaseRetrievalRandom2012}. More recent proposals include semidefinite programming \cite{candesPhaseLiftExactStable2013, waldspurgerPhaseRecoveryMaxCut2015}, matrix completion \cite{candesPhaseRetrievalMatrix2013}, and an alternating-projection method called \emph{Relaxed-Reflect-Reflect} (RRR) \cite{elserComplexityBitRetrieval2018}. Although RRR lacks a comprehensive theoretical foundation, subsequent investigations \cite{levinNoteDouglasRachfordGradients2020} reported robust empirical performance. In particular, \cite{elserBenchmarkProblemsPhase2018} demonstrated that RRR outperforms both the “Phase Lift” approach of \cite{candesPhaseLiftExactStable2013} and Wirtinger flow methods \cite{chenSolvingRandomQuadratic2017} in a variety of benchmark tests.

In this work, we demonstrate that the Coded Diffraction Pattern (CDP) variant of phase retrieval can be reformulated as an \textit{XY} Hamiltonian minimization problem, amenable to physical solvers that emulate the same Hamiltonian, for instance, through gain-based oscillator networks~\cite{cummins2025ising, syedPhysicsEnhancedBifurcationOptimisers2022b}. Under moderate noise levels in the measured amplitudes, we show numerically that a gain-based system can outperform the state-of-the-art Relaxed-Reflect-Reflect (RRR) algorithm and reliably reconstruct complex-valued experimental data with high accuracy.

\section{Methods}
\subsection{Mapping Phase Retrieval into XY Problem}
The work of \cite{waldspurgerPhaseRecoveryMaxCut2015} first recast phase retrieval as a non-convex quadratic program, facilitating its solution via a suitable relaxation method. Following a similar line of reasoning, we show how to reformulate phase retrieval in the CDP experiment setting as an \textit{XY} Hamiltonian minimization problem.

We start by defining the unknown phase of the observation as $\vect{p}$, where all elements of this vector are complex and have unit amplitude i.e. $p_i \in \{e^{i\theta_i} | \theta_i \in [0, 2\pi)\}$.
By this definition, we have:
\begin{equation}
    \sum_{j=1}^N{A_{ij}x_j}=b_i p_i \quad \text{for all $i=1,...,M$} .
    \label{eqn:complete_observation}
\end{equation}
Suppose that the complete observation information including the phase information $\vect{p}$ and the amplitude information $\vect{b}$ is known, then the process of trying to find $\vect{x}$ that satisfy the constraint given by Eq.~\eqref{eqn:complete_observation} is equivalent to the minimisation of the cost function $\mathscr{E}$ given by:
\begin{equation}
    \mathscr{E}(\{x_j\}) = \sum_{i=1}^M \left|{\sum_{j=1}^N A_{ij}x_j - b_i p_i}\right|^2,
\end{equation}
where the minimisation is over all possible sets of $\{x_j\}$ for $j \in \{1, \cdots, N\}$.
If the observations $\vect{b}$ and $\vect{p}$ are exact, the minimum value of this expression should be $0$, but if uncertainty exists in either of them, then this expression is a least square problem over $\vect{x}$.
The solution to the least square problem is given by the Moore-Penrose inverse, also known as the pseudoinverse of $\matr{A}$, denoted as $\matr{A}^{\dagger}$:
\begin{equation}
    x_i=\sum_j^N{A^{\dagger}_{ij} b_j p_j}
    \label{eqn:least_sq_sol}
\end{equation}
Hence, for any given set of observation phase $\{p_i\}$, the minimal cost function $\mathscr{E}$ is given by:
\begin{equation}
    \mathscr{E}(\{p_i\}) = \sum_i^M\left|{\sum_{j}^{N}\sum_{k}^{M}{A_{ij}A^{\dagger}_{jk} b_k p_k} - b_i p_i}\right|^2
    \label{eqn:optimise_over_p}
\end{equation}
This means that to solve the original phase retrieval problem, one needs to find a set of $\{p_i\}$ that minimises this cost function $L$, which is essentially a QCO problem over a set of variables $\{p_i\}$ that all have unit amplitude and unconstrained phases, so this problem can be mapped into an $XY$ Hamiltonian minimisation problem.

To put it into an explicit $XY$ Hamiltonian form, we rearrange Eq.~\eqref{eqn:optimise_over_p} as follows:
\begin{eqnarray}
    \mathscr{E}(\{p_i\}) &&= \sum_i^M\left|{\sum_{j,k}^{N,M}{A_{ij}A^{\dagger}_{jk} b_k p_k} - \sum_k^M {\delta_{ik} b_k p_k}}\right|^2 \nonumber\\
    &&= \sum_i^M \left| \sum_k^M {\left( \sum_j^N A_{ij} A^{\dagger}_{jk} - \delta_{ik} \right)b_k p_k} \right|^2 \nonumber\\
    &&=\sum_i^M \left| \sum_k^M G_{ik} b_k p_k \right|^2 \nonumber \\
    &&=\sum_{i}^M \sum_{jk}^M {G_{ij} G_{ki}^* b_j p_j b_k p_k^*} \nonumber \\
    &&=-\sum_{jk}^M \tilde{J}_{jk} p_j p_k^*,
    \label{eqn:raw_pr_xy_hamiltonian}
\end{eqnarray}
where from the second line to the third line for ease of notation we defined $G_{ik}=\sum_j^N A_{ij}A^{\dagger}_{jk}-\delta_{ik}$, and from the fourth line to the last we identified the coupling matrix elements of the equivalent XY Hamiltonian to be $\tilde{J}_{jk}=-\sum_i^M{G_{ij} G_{ki}^* b_j b_k}$.

In principle, Eq.~\eqref{eqn:raw_pr_xy_hamiltonian} is already in the form of an XY Hamiltonian we first introduced in Eq.~\eqref{eqn:xy_hamiltonian}. 
Noting that $\matr{A} \matr{A}^{\dagger}$ is Hermitian and $\matr{A} \matr{A}^{\dagger}\matr{A} \matr{A}^{\dagger} = \matr{A} \matr{A}^{\dagger}$ by property of the pseudoinverse, one can simplify the expression for $\tilde{J}_{ij}$ to:
\begin{equation}
    \tilde{J}_{ij} = \left( \sum_k^N {A_{ik} A_{kj}^{\dagger}} - \delta_{ij} \right)b_i b_j,
    \label{eqn:jtilde}
\end{equation}
which leads to an equivalent XY Hamiltonian:
\begin{eqnarray}
    \mathscr{E}(\{p_i\}) &&= -\sum_{ij}^M \left( b_i b_j \sum_k^N A_{ik} A_{kj}^{\dagger} \right)p_i p_j^* + \sum_{ij}^M \delta_{ij} p_i p_j^* \nonumber \\
    && = H_{XY}(\{p_i\}) + M,
\end{eqnarray}
where $M$ is as before  the dimension of observation vector $\vect{b}$.
Hence, minimising $\mathscr{E}$ is equivalent to minimising the simpler XY Hamiltonian $H$, whose coupling matrix elements are given by:
\begin{equation}
    J_{ij}=\sum_k^N A_{ik} A_{kj}^{\dagger} b_i b_j.
\end{equation}
This coupling matrix was then used as input to the simulated gain-based system, with outcomes  presented in the Results section.

We note that the Moore--Penrose pseudoinverse $A^{\dagger}$ may become ill-conditioned when $\operatorname{rank}(A) < N$.  
A Tikhonov-regularised alternative,
\[
(c\,\mathbf I + A A^{\ast})^{-1} A^{\ast},
\]
with $c>0$, can be used to stabilise the inversion at the cost of losing the compact expression in Eq.\,(\ref{eqn:jtilde}).  
In practice, choosing $c \sim 10^{-3}\lVert A\rVert_2^{\,2}$ leaves all numerical results within the error bars reported below.

\subsection{Simulated Gain-based System}
To minimise the $XY$ Hamiltonian $H_{XY}$ for a given phase retrieval problem, we simulate an oscillator network following the gain-based dynamics given by
\begin{eqnarray}
    \frac{d \psi_i}{dt} &&= \left(\gamma_i - |\psi_i|^2 \right) \psi_i + \sum_j^M J_{ij} \psi_j
    \label{eqn:gd_dynamics} \\
    \frac{d \gamma_i}{dt} &&= \epsilon \left(1 - |\psi_i|^2 \right),
    \label{eqn:feedback_dynamics}
\end{eqnarray}
where $\gamma_i \in \mathbb{R}$ is the effective injection rate of oscillator $i$ (gain minus losses), $\psi_i \in \mathbb{C}$ characterise each oscillator, and $J_{ij}$ specifies the coupling strength between oscillators.
$J_{ij}$ is calculated from the given phase retrieval problem according to our previous discussion.
$\epsilon$ is an externally controlled positive constant, which measures the responsiveness of the gain of each oscillator to the amplitude variations of each oscillator. Eqs.(\ref{eqn:gd_dynamics})–(\ref{eqn:feedback_dynamics}) faithfully reproduce the gain–dissipative evolution observed in networks of exciton-polaritons hosted in semiconductor micro-cavities \cite{berloff2017realizing}, in coupled-laser arrays \cite{nixon2013observing},  in  driven photonic oscillator lattices \cite{toebes2022dispersive}, and spin wave Ising machines \cite{litvinenko2023spinwave}. Hardware platforms that do not possess intrinsic gain control, most notably the spatial photonic Ising machine (SPIM) \cite{pierangeli_large-scale_2019, veraldi2025fully}, can nevertheless implement the same update rule by applying digital feedback to the spatial-light modulator after each optical pass, thereby emulating the gain–loss loop in silico while retaining full optical parallelism.
This kind of dynamics was formulated and  studied in \cite{kalinin2018networks, kalininGlobalOptimizationSpin2018, syedPhysicsEnhancedBifurcationOptimisers2022b}, showing that the dynamics of this system leads to stationary states  close to or at the global minimum of the XY Hamiltonian specified by the coupling matrix $\matr{J}$ with high probability.
Equation~\eqref{eqn:gd_dynamics} encapsulates the main dynamics of $\psi_i$ with the interplay of the effective gain, non-linear loss, and the coupling terms.
Eq.~\eqref{eqn:feedback_dynamics} provides a feedback mechanism that pushes the amplitudes of all oscillators towards $1$. 
It was previously reported that this feedback mechanism is crucial for the gain-dissipative dynamics to produce good solutions close to the true global minimum of the XY Hamiltonian \cite{kalinin2018networks}. 

In principle, the gain-based optimiser works as follows.
At the start of the dynamical evolution, the oscillator network has a set of highly negative effective gains $\gamma_i$, so the system has a stable fixed point at $\psi_i=0$.
Due to the gain dynamics given by Eq.~\eqref{eqn:feedback_dynamics}, the gains $\gamma_i$ increase over time and eventually cross a critical value at which supercritical Hopf bifurcation occurs \cite{kalininGlobalOptimizationSpin2018, syedPhysicsEnhancedBifurcationOptimisers2022b}.
The $\psi_i=0$ fixed point becomes unstable, and oscillators spontaneously increase to some non-zero amplitudes and start to have well-defined phases.
Over time, all amplitudes $|\psi_i|$ approaches $1$, while the phases of oscillators also approach their stationary values, which will be read out as the solution to the XY problem.

In our simulations, we first initialised the amplitude of $\psi_i$ to some random small but non-vanishing values uniformly distributed in the range $(0, 0.1)$ and initialised their phase uniformly randomly in the range $[0, 2\pi)$.
Initial gains $\gamma_i$ were initialised uniformly randomly in such range of values, so that initial stages of the evolution are below the threshold, which means that oscillator amplitudes will tend to decay to $0$ if gains are held constant at this level.
The system was then evolved until it reached a stationary state, and the phases of oscillators were used as the spin configurations $\vect{p}$ for the $XY$ problem.
This can then be substituted into Eq.~\eqref{eqn:least_sq_sol} to find the solution $\vect{x}$ to the original phase retrieval problem.

\subsection{Generation of CDP Phase Retrieval Problems}
For a given sample vector $\vect{x} \in \mathbb{C}^N$, we had to generate observation vector $\vect{b} \in \mathbb{R}^M$ under CDP experiment framework to serve as the input to our $XY$ minimiser.
A set of $L$ random filters was first generated.
Each filter could be represented as a diagonal matrix $\matr{D}^{(i)}$ where $i=1,...,L$, and each diagonal elements $D^{(i)}_{jj}$ was uniformly randomly selected from $\{1, i, -1, -i\}$, corresponding to a phase shift of $0$, $\pi/2$, $\pi$, $3\pi/2$ respectively.
These diagonal matrices were stacked along the rows to produce a $M \times N$ matrix $\matr{G}$, where $M=NL$, as follows:
\begin{equation}
    \matr{G} =
    \begin{pmatrix}
         \matr{D}^{(1)} \\
         \matr{D}^{(2)} \\
         \vdots \\
         \matr{D}^{(L)}
    \end{pmatrix}.
\end{equation}

Similarly, we represent the action of the Fourier transform on the sample vector by using the discrete Fourier transform (DFT) matrix defined as:
\begin{equation*}
\mathbf{F}_N = \frac{1}{\sqrt{N}}
    \begin{pmatrix} 
    1 & 1 & 1 & \cdots & 1 \\
    1 & \omega_N & \omega_N^2 & \cdots & \omega_N^{N-1} \\
    1 & \omega_N^2 & \omega_N^4 & \cdots & \omega_N^{2(N-1)} \\
    \vdots & \vdots & \vdots & \ddots & \vdots \\
    1 & \omega_N^{N-1} & \omega_N^{2(N-1)} & \cdots & \omega_N^{(N-1)^2}
    \end{pmatrix},
\end{equation*}
where $\omega_N=e^{-2\pi i/N}$ is the N-th root of unity.
We construct an \(M \times N\) matrix \(\mathbf{F}\) by stacking \(L\) copies of the DFT matrix ${\bf F}_N$ as consecutive row blocks, effectively placing each DFT matrix on top of the next
\begin{equation}
    \matr{F} =
    \begin{pmatrix}
         \matr{F}_N \\
         \vdots \\
         \matr{F}_N
    \end{pmatrix}.
    \label{eqn:combined_dft_matrix}
\end{equation}

Then the elements of the  matrix $\matr{A}$  are defined by as
\begin{eqnarray}
    A_{ij} = F_{ij} G_{ij},
\end{eqnarray} 
while the observation vector $\vect{b}$ can be obtained by substituting matrix $\matr{A}$ and $\vect{x}$ into Eq.~\eqref{eqn:phase_retrieval}.
The observation vector $\vect{b}$ and the generated matrix $\matr{A}$ are supplied to our gain-based optimiser Eq.~(\ref{eqn:gd_dynamics})-(\ref{eqn:feedback_dynamics}) and the phases $\theta_i$ are found as the stationary states. The sample vector $\vect{x}$ is then reconstructed from  Eq.~(\ref{eqn:least_sq_sol}).

The block stacking of all Fourier operators ${\bf F}$ 
and diagonal masks ${\bf D}^{(i)}$
 is mathematically identical to the decomposition of a fully-connected interaction matrix into multiple Mattis subproblems recently realised in a fully programmable SPIM via focal-plane division, where the energies of all sub-Hamiltonians are computed in parallel on distinct camera regions \cite{veraldi2025fully}. This analogy suggests that an experimental CDP phase-retrieval setup could exploit the same optical parallelism, processing the 
$L$
 masks in a single physical shot and obtaining an 
$L$-fold reduction in acquisition.

In this study, we also considered the case where $\vect{b}$ is noisy, which is to be expected in realistic experimental data.
In this case, a noisy observation vector $\vect{\tilde{b}}$ is produced by adding a normally distributed random noise to each element of the noiseless observation vector $\vect{b}$, i.e.  $\tilde{b}_i= b_i + \xi_i$ with $\xi_i \sim \mathcal{N}(0, \sigma^2)$.
The variance of noise $\sigma$ was used to control the magnitude of noise in the observational data so that we could investigate its impact on the performance of the gain-based system in solving the phase retrieval problem.
To quantify the amount of noise in the given noisy observation vector, we define the signal-to-noise ratio (SNR) as follows:
\begin{equation}
    \text{SNR}=10 \log_{10} \frac{\lVert \vect{b} \rVert_2}{\lVert \vect{\tilde{b}} - \vect{b} \rVert_2},
\end{equation}
which is conventionally measured in a logarithmic scale and quoted in unit of decibel. 
In this expression, $\lVert \cdot \rVert_2$ denotes vector 2-norm.

\subsection{Performance Evaluation}
To measure the quality of the calculated solution, we used metrics that were also used previously in \cite{elserBenchmarkProblemsPhase2018, waldspurgerPhaseRecoveryMaxCut2015}.
The most direct metric to measure the success of the phase retrieval algorithm is the Euclidean distance between the observation vector calculated from the recovered sample vector $\vect{\tilde{x}}$ and the given observation vector $\vect{b}$, normalised over the vector norm of the observation vector $\vect{b}$.

For clearer visual representation, it is more convenient to express this quantity on a logarithmic scale, similar to SNR. Therefore, we define the relative observation error (ROE) in decibels as:
\begin{equation}
    \text{ROE} = 10\log_{10} \frac{\lVert |\matr{A} \vect{\tilde{x}}| - \vect{b} \rVert_2}{\lVert \vect{b} \rVert_2},
\end{equation}
where $| \cdot |$ denotes taking the amplitude element-wise, and $\vect{\tilde{x}}$ is the sample vector calculated by the phase retrieval algorithm.
In experiments, the true sample vector $\vect{x}$ is typically unavailable, so the ROE would be an appropriate way to evaluate the quality of the recovered solution.
However, in generated datasets where the true sample vector $\vect{x}$ is known a priori, we can instead measure the Euclidean distance between the recovered sample vector $\vect{\tilde{x}}$ and the true sample vector $\vect{x}$.
Note that the recovered sample vector $\vect{\tilde{x}}$ may have a global phase shift relative to $\vect{x}$ while reproducing the exact same observation vector $\vect{b}$.
Hence, the error metric should be defined as the minimum Euclidean distance between $e^{i\theta}\vect{\tilde{x}}$ and $\vect{x}$ for all $\theta \in [0, 2\pi)$.
This quantity is also expressed in decibels and is referred to as the relative sample error (RSE), which is defined as:
\begin{equation}
    \text{RSE} = 10\log_{10} \left( \min_{\theta} \frac{\lVert e^{i\theta} \vect{\tilde{x}} - \vect{x} \rVert_2}{\lVert \vect{x} \rVert_2} \right).
    \label{eqn:sample_error_def}
\end{equation}
In a phase retrieval problem, this error is usually the most important quantity because it measures how close the recovered solution is to the original sample.

\begin{figure}
    \centering
    \includegraphics[width=\columnwidth]{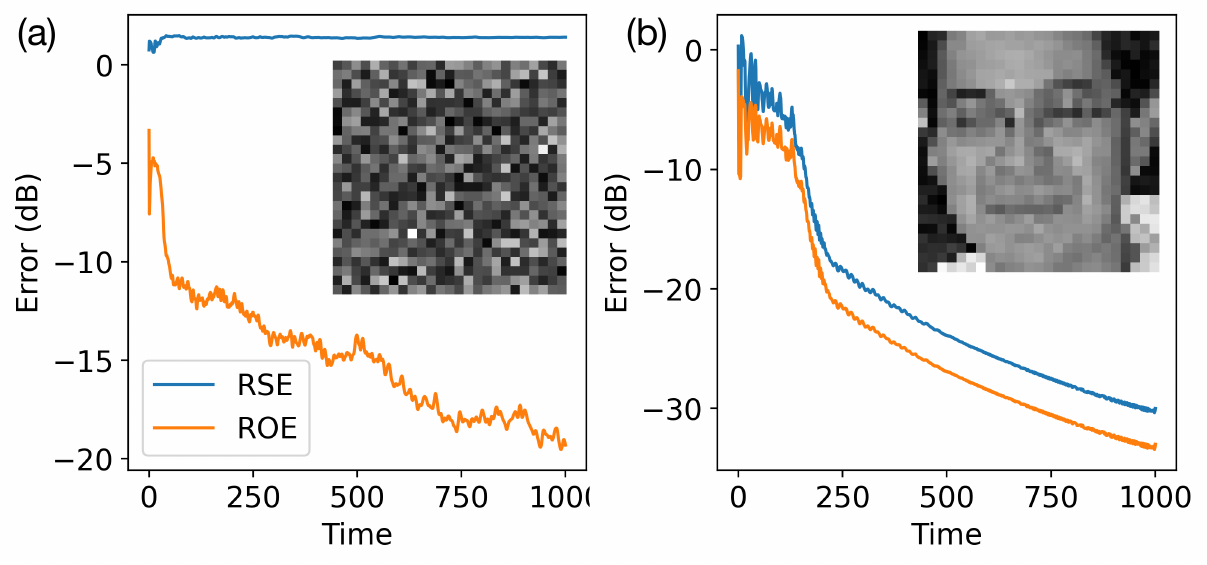}
    \caption{ Comparison of phase retrieval problems with different number of phase masks.
    (a) Time evolution of errors for the phase retrieval problem under CDP experiment framework with 2 phase masks.
    The inset gives the final reconstructed image produced by the phase retrieval algorithm.
    (b) Time evolution of errors for the phase retrieval problems with 5 phase masks.
    In both cases the problems were solved with the gain-dissipative system with a random initial condition.
    The original image, which is visually identical to the image shown in inset of (b), is from the labelled face in the wild (LFW) dataset. 
    }
    \label{fig:1_mask_no}
\end{figure}

The two error metrics are closely related but not equivalent, and both are required to give a complete picture about the well-posedness of the phase retrieval problem itself and the performance of the phase retrieval algorithm used.
For example, if a solution calculated by a phase retrieval algorithm yields a small ROE value but a large RSE value, this indicates that the algorithm is performing well, but the problem itself is poorly defined because it has more than one degenerate ground states (i.e. more than one $\vect{x}$ can all produce the same observation vector $\vect{b}$).
This is because the phase retrieval algorithm only has access to $\vect{b}$ and $\matr{A}$, so any solution $\vect{\tilde{x}}$ that can minimise $\lVert |\matr{A} \vect{\tilde{x}} | - \vect{b} \rVert_2$ are equally good to the algorithm, even if it might be far from the true $\vect{x}$ from which the problem was first constructed.
This situation is illustrated in Fig.~\ref{fig:1_mask_no}(a) and (b) where the simulated gain-based system tries to recover an image from an observation vector produced by 2 phase filters.
While ROE keeps decreasing, RSE has remained largely flat and the system failed to recover the original image.
As far as the phase retrieval method is concerned, it is performing well because it is able to find a vector $\vect{\tilde{x}}$ that produces an observation vector $\vect{\tilde{b}}$ that is very close to the known observation vector $\vect{b}$.
This means that this phase retrieval problem under the CDP experiment framework with only 2 phase masks is ill-defined, because it has degenerate ground states.

When both RSE and ROE are small, it suggests that the phase retrieval problem is well defined and the solution found is close to the true solution.
This case is shown in Fig.~\ref{fig:1_mask_no}(c) and (d) where the same gain-dissipative system recovers the same image from an observation vector produced by 5 phase filters.
RSE and ROE decrease in tandem, indicating that the algorithm is approaching the planted ground state.
Hence, when studying phase retrieval algorithms, it is crucial to consider both the ROE and RSE to determine whether the phase retrieval method itself or the problem at hand is responsible for the failure to recover the original signal.

\subsection{Comparison with Existing Algorithm}
To compare the gain-dissipative system with established phase retrieval methods, we focus on the RRR algoritm for phase retrieval ~\cite{elserComplexityBitRetrieval2018}. Originally designed for the sparse variant of phase retrieval, RRR belongs to the family of alternating projection methods, similar to classic techniques such as the Gerchberg--Saxton (GS) method~\cite{gerchbergHolographyFringesElectron1972} and Fienup’s hybrid input--output (HIO) scheme~\cite{fienupPhaseRetrievalAlgorithms1982}. These algorithms operate in an iterative fashion, applying two distinct projection operators in sequence at each iteration. Among them, RRR has shown particularly strong empirical performance, surpassing multiple modern phase retrieval methods~\cite{elserBenchmarkProblemsPhase2018}.

The RRR algorithm  starts with a guessed observation vector $\vect{b}_0$, and employs two projections $P_1$ and $P_2$. 
When given a (generally complex) estimated observation vector $\vect{b}_n$, $P_1$ keeps only the $S$  largest elements in the vector and sets other to 0, where $S$ is the given sparsity constraint in the observation data;
For $P_2$, when given $\vect{b}_n \in \mathbb{C}^M$, it keeps the phase of each elements but overwrites their amplitudes with the known correct amplitudes in $\vect{b} \in \mathbb{R}^M$.
Overall, in each iteration, the algorithm applies $P_1$ and $P_2$ as follows:
\begin{equation}
    \vect{b_{n+1}} = \vect{b_n} + \beta \left[ P_2\left( 2P_1(\vect{b_n}) -\vect{b_n} \right) - P_1(\vect{b_n}) \right],
    \label{eqn:rrr_iteration}
\end{equation}
where $\beta$ is a constant parameter.
We found the value $\beta=0.5$ used by authors of \cite{elserBenchmarkProblemsPhase2018} generally produced good results.

To adapt RRR for the CDP variant of phase retrieval problem, we follow the method proposed in \cite{elserBenchmarkProblemsPhase2018} and modify projection $P_1$ to the following:
\begin{equation}
    P_1(\vect{b}_n)=\matr{A} \matr{A}^{\dagger} \vect{b}_n.
    \label{eqn:modified_projection}
\end{equation}
One can motivate this projection by considering that $\matr{A}^{\dagger} \vect{b_n}$ makes use of all $L$ sets of observations, unique to the CDP formulation, to produce an ``average'' estimated sample vector $\vect{x}_n$ based on all available observations, and then using this best estimation to produce the next $\vect{b}_{n+1}$ by calculating $\matr{A} \vect{x}_n$, which in turn ensures that $\vect{b}_{n+1}$ produced this way remains in the range of $\matr{A}$.
We then substituted this modified projection \(P_1\) into the iterative scheme of Eq.~\eqref{eqn:rrr_iteration}, keeping the original \(P_2\) operator and \(\beta=0.5\), and applied the resulting RRR method to the same phase retrieval instances used by the gain-based system.

The classical GS method, which was one of the first proposed heuristic method for solving phase retrieval problems, used the same $P_1$ and $P_2$ projections, but a simpler iterative formula:
\begin{equation}
    \vect{b}_{n+1} = P_2(P_1(\vect{b}_n)).
    \label{eqn:fienup_iteration}
\end{equation}
This iterative heuristic can also be applied to CDP phase retrieval problems with modified projection $P_2$ given by Eq.~\eqref{eqn:modified_projection}.
The quality of solutions found by these methods were measured by the performance metrics presented in Performance Evaluation section and the findings are presented in the Results section.

\section{Results}
\label{sec:results}
\subsection{Performance with Real-valued Sample Vectors}

\begin{figure}
    \centering
    \includegraphics[width=\columnwidth]{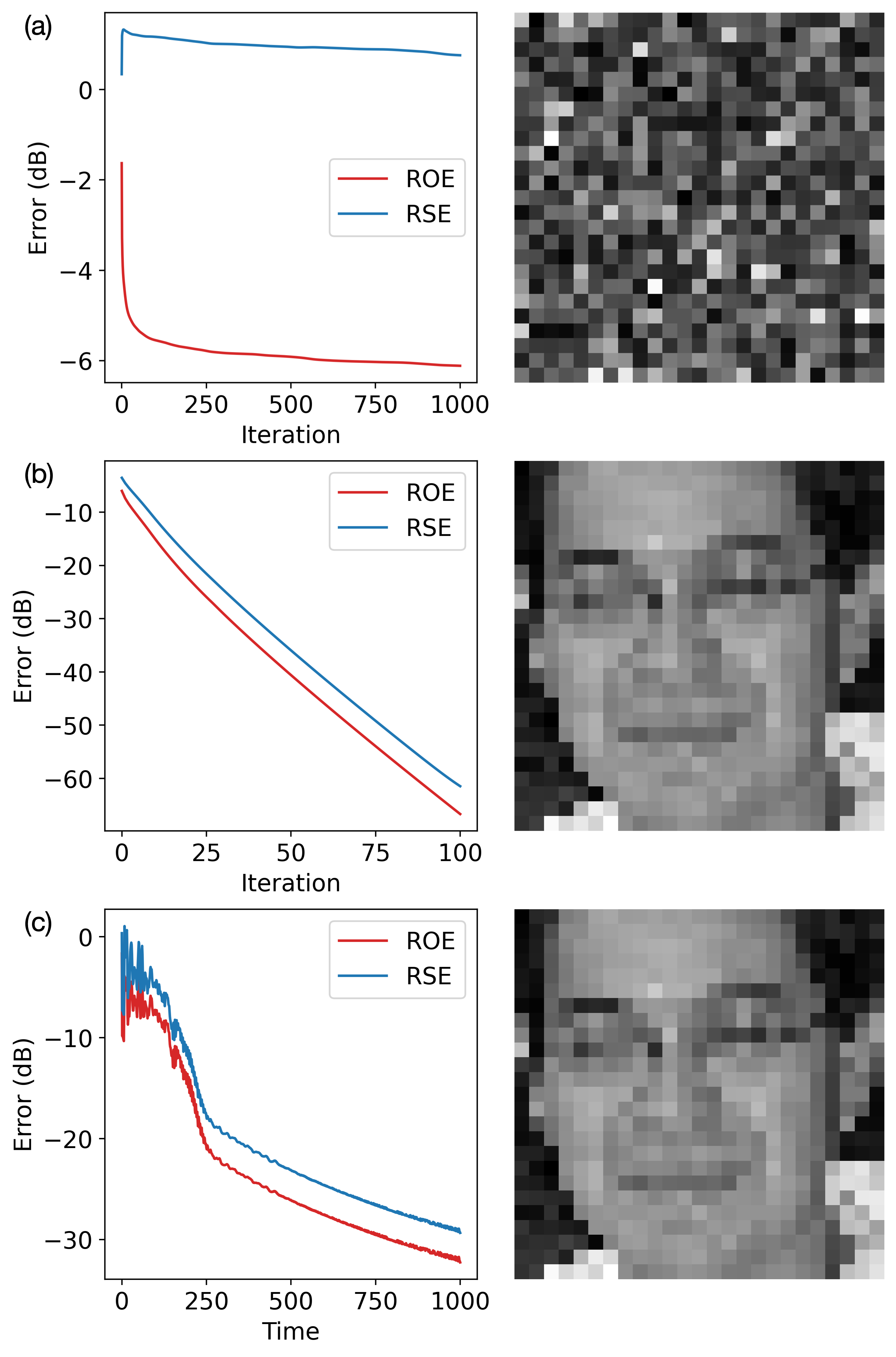}
    \caption{Comparison of GS method and gain-based system performance in recovering a real-valued image.
    (a) Error evolution and final reconstructed image with GS method starting from a complex-valued random initial condition whose phase is uniformly randomly distributed in the range $[0, 2\pi)$ and whose amplitude is the known observation vector $\vect{b}$.
    (b) Error evolution and final reconstructed image with GS method starting from an initial condition whose phase is obtained by multiplying $\matr{A}$ with a random real-valued vector $\vect{\tilde{x}}$, and whose amplitude is the known observation vector $\vect{b}$.
    (c) Error evolution and final reconstructed image with gain-based method starting from the exact same initial condition as used in (a).
    }
    \label{fig:2_real_img_compare}
\end{figure}
Many studies~\cite{latychevskaiaIterativePhaseRetrieval2018, elserBenchmarkProblemsPhase2018, shechtmanPhaseRetrievalApplication2015} have evaluated phase retrieval algorithms using images that are real and strictly positive, represented by two-dimensional matrices. Although such images offer a convenient test bed, they restrict \(\mathbf{x}\) to real values only, whereas real-world diffraction data are generally complex. Nevertheless, these simple test cases already reveal key limitations of traditional phase retrieval algorithms.

For instance, we used the positive real-valued image depicted in Fig.~\ref{fig:1_mask_no} to construct a CDP-based phase retrieval problem with five phase filters. The resulting observation vector \(\mathbf{b}\) was then presented to different solvers. As shown in Fig.~\ref{fig:2_real_img_compare}(a), when GS method began from a random complex initialization for \(\mathbf{b}\), it quickly became stuck in a local minimum after about 1000 iterations, failing to reproduce any recognizable features of the original image. By contrast, starting from the same initial condition, the gain-dissipative system (Fig.~\ref{fig:2_real_img_compare}(c)) followed a markedly different trajectory and produced significantly lower errors, ultimately reconstructing the image with high fidelity.

We further observed that a priori knowledge of \(\mathbf{x}\) being real and positive can substantially simplify phase retrieval. For example, by multiplying \(\mathbf{A}\) with a random positive real-valued vector \(\mathbf{\tilde{x}}\) and extracting its phase as the initial guess for GS, we obtained the error evolution and final reconstruction depicted in Fig.~\ref{fig:2_real_img_compare}(b). This carefully chosen initialization, which already exhibits a lower error than a random complex guess, allowed the GS method to recover the underlying image. Clearly, such information (i.e., that \(\mathbf{x}\) is non-negative real) makes the phase retrieval problem more tractable. However, real-world experimental data generally produce complex-valued \(\mathbf{x}\) without providing a straightforward initialization. Consequently, the GS method often encounters difficulty in practical scenarios. In the following section, we therefore focus on more general, complex-valued samples and benchmark the gain-dissipative solver in that setting.

\begin{figure}
    \centering
    \includegraphics[width=\columnwidth]{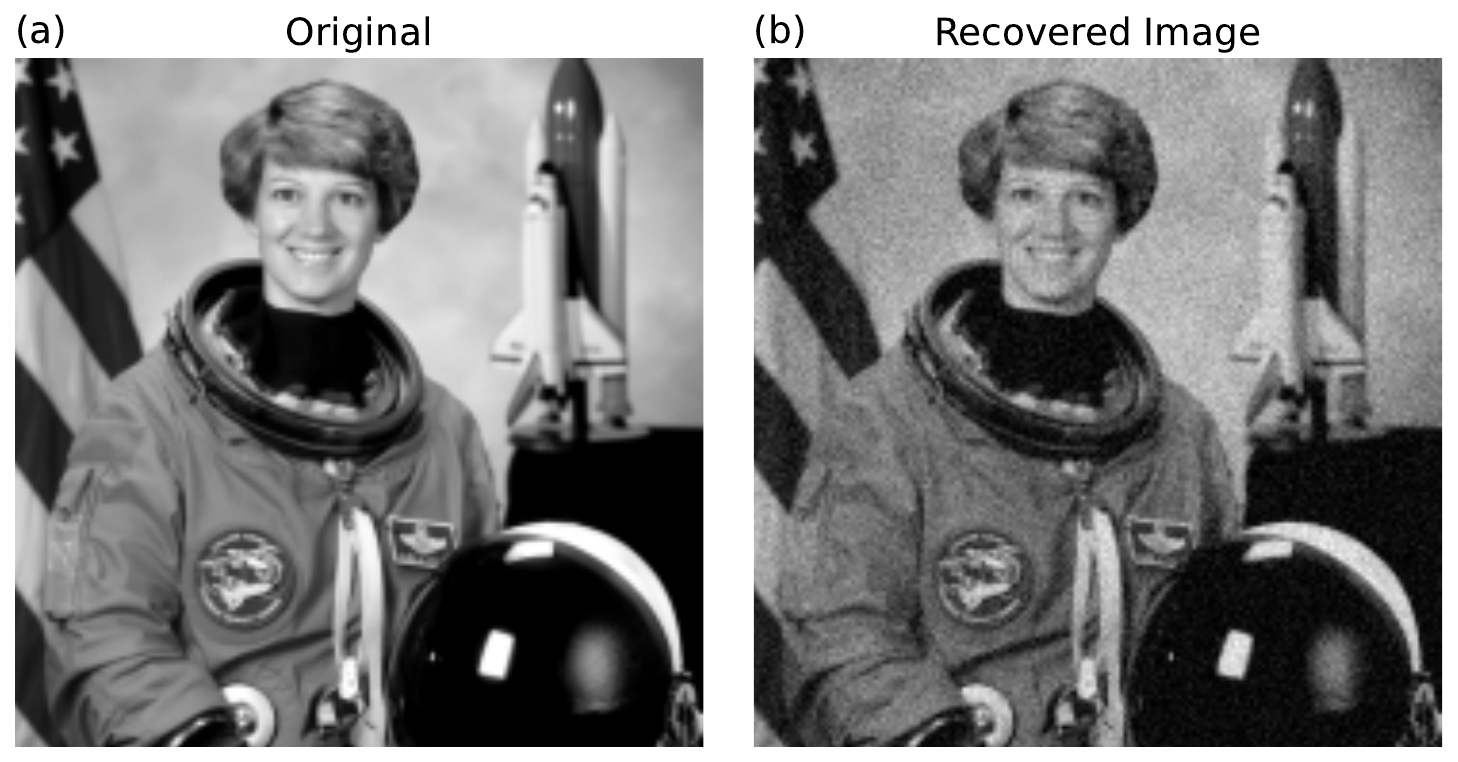}
    \caption{
    Phase retrieval with a large-scale sample vector using the gain-based system.
    (a) A \(180 \times 180\) pixel grayscale image of an astronaut (sourced from NASA’s Great Images Database, public domain). This image is used as the sample vector in a CDP-based phase retrieval setup with 8 phase filters, yielding an observation vector \(\mathbf{b}\) of length \(259{,}200\).
    (b) The final reconstruction after the gain-based system evolves for \(t=5\) from a random initial condition. The resulting RSE is \(-9.4\) and the ROE is \(-12\).
    } 
    \label{fig:2_1_real_large_img}
\end{figure}

As a final demonstration using real-valued data, Fig.~\ref{fig:2_1_real_large_img}(a) shows a high-resolution grayscale image of size \(180 \times 180\), totaling 32\,400 real-valued pixels. Despite this large problem dimension, the gain-based method successfully reconstructs the main features of the image after a short simulation (duration \(t=5\)), as shown in Fig.~\ref{fig:2_1_real_large_img}(b). Notably, this was achieved without leveraging the fact that the target sample vector is purely real. While the final RSE remains moderately high at \(-8.5\), the essential image structure is clearly recognizable, albeit with visible background noise.

Although the gain-based method easily solves real-valued phase retrieval problems, most practical applications involve recovering \emph{complex}-valued data. Consequently, the remaining sections focus on benchmark cases with complex-valued samples to more accurately reflect real-world experimental conditions.

\subsection{Performance with Random Complex-valued Sample Vectors}
\begin{figure}
    \centering
    \includegraphics[width=\columnwidth]{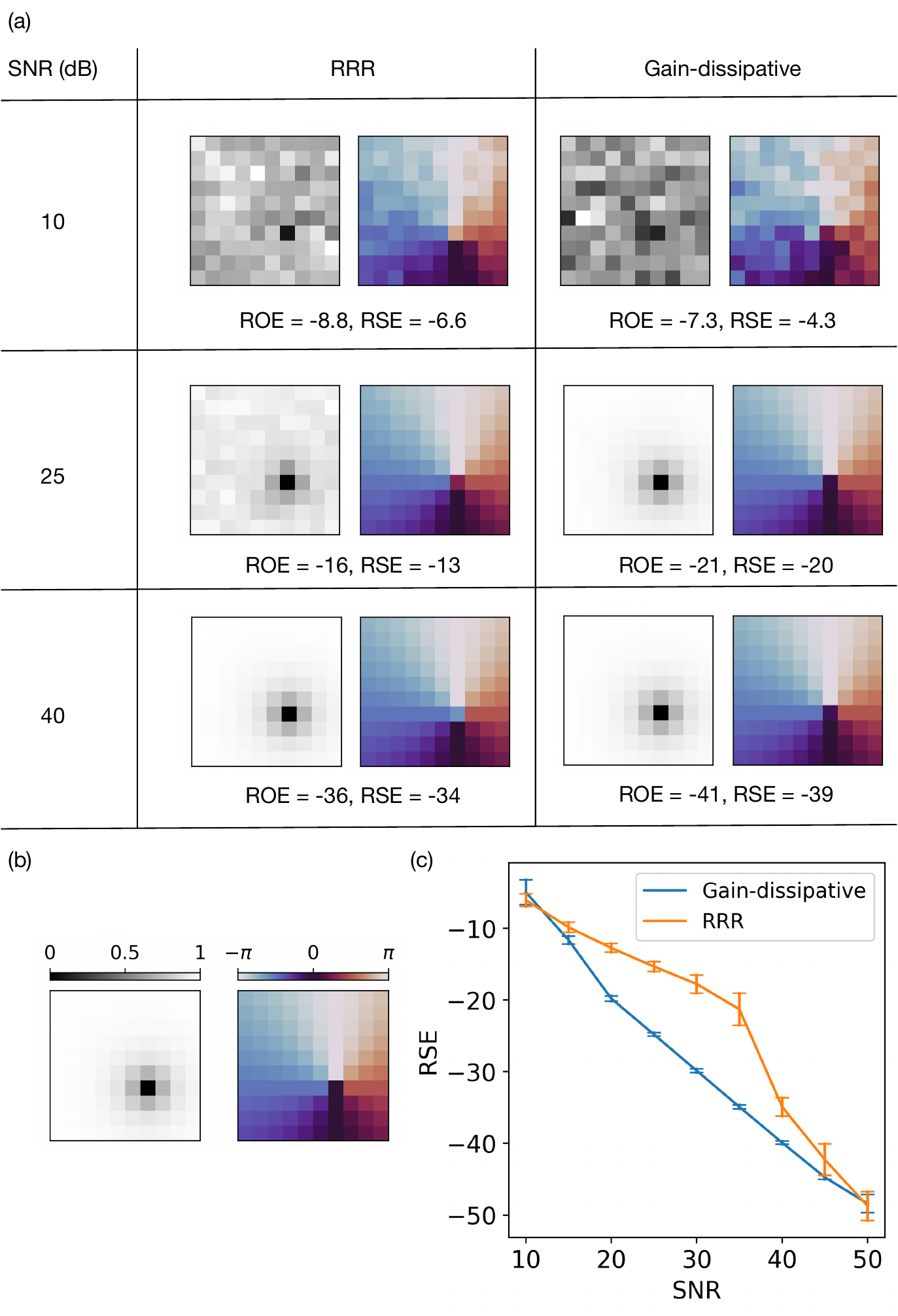}
    \caption{Phase retrieval of a two-dimensional vortex in the presence of noise, comparing RRR and the gain-based system.
    (a) Each panel displays the reconstructed sample vector \(\mathbf{\tilde{x}}\), where the grayscale image encodes amplitude and the  color image encodes phase. For each noise level, both RRR and the gain-based method start from the same initial condition. Here, RRR runs for 10{,}000 iterations, while the gain-based system is evolved to \(t=1{,}000\).
    (b) The ground-truth sample vector \(\mathbf{x}\) that describes a 2D vortex, showing amplitude (left) and phase (right).
    (c) Phase retrieval error (RSE) versus the signal-to-noise ratio (SNR). Each data point represents the average of 20 random instances, where the vortex core is placed at different positions and each algorithm is initialized randomly.  Error bars denote the standard deviation of the final RSE values.
    }
    \label{fig:3_complex_image_visual_compare}
\end{figure}

We begin by investigating the reconstruction of a complex-valued image that can potentially be produced in an experiment: a two-dimensional vortex, where the vortex flow is given by the gradient of phase at each point in a plane.
The field value at each point can be approximated by:
\begin{equation}
    v(x,y)=\frac{(x-x_0)+i(y-y_0)}{\sqrt{(x - x_0)^2+(y-y_0)^2+\xi^2}},
\end{equation}
where $(x_0, y_0)$ is the centre of the vortex and $\xi$ is the size of the vortex core \cite{berloff2004pade}.
The amplitude, $|v|$ (greyscale) and phase of the vortex $\arg(v)$  (color) can be visualised as shown in Fig.~\ref{fig:3_complex_image_visual_compare}(b).

In most practical experiments, the observed amplitudes \(\mathbf{\tilde{b}}\) deviate from the ideal measurements \(\mathbf{b}\) due to noise. As illustrated in Fig.~\ref{fig:3_complex_image_visual_compare}(a), both the RRR and gain-based solvers were presented with noisy observations of varying magnitude. In the high-noise regime (small SNR values), neither method recovered a solution close to the original sample \(\mathbf{x}\), shown in Fig.~\ref{fig:3_complex_image_visual_compare}(b). Under moderate noise (SNR \(\approx 20\)), the RRR reconstruction exhibited pronounced amplitude distortions, whereas the gain-based solver produced a visually accurate approximation of \(\mathbf{x}\). This improvement is reflected quantitatively by the lower RSE values achieved by the gain-dissipative method, a performance gap that persists until around \(\text{SNR} \approx 40\). Above this threshold, the visual differences between the two solvers diminish, although residual discrepancies in ROE and RSE remain.
Figure~\ref{fig:3_complex_image_visual_compare}(c) summarizes the impact of noise on reconstruction accuracy, revealing that the gain-based approach and RRR perform similarly in the high-noise (\(\text{SNR} < 10\)) and low-noise (\(\text{SNR} > 40\)) regimes. Notably, however, the gain-dissipative solver holds a distinct advantage in the intermediate range \(10 < \text{SNR} < 40\), where the difference in RSE is both statistically and visually significant. In these conditions, the gain-dissipative system often recovers important structural details that RRR loses in noise.

\begin{figure}
    \centering
    \includegraphics[width=\columnwidth]{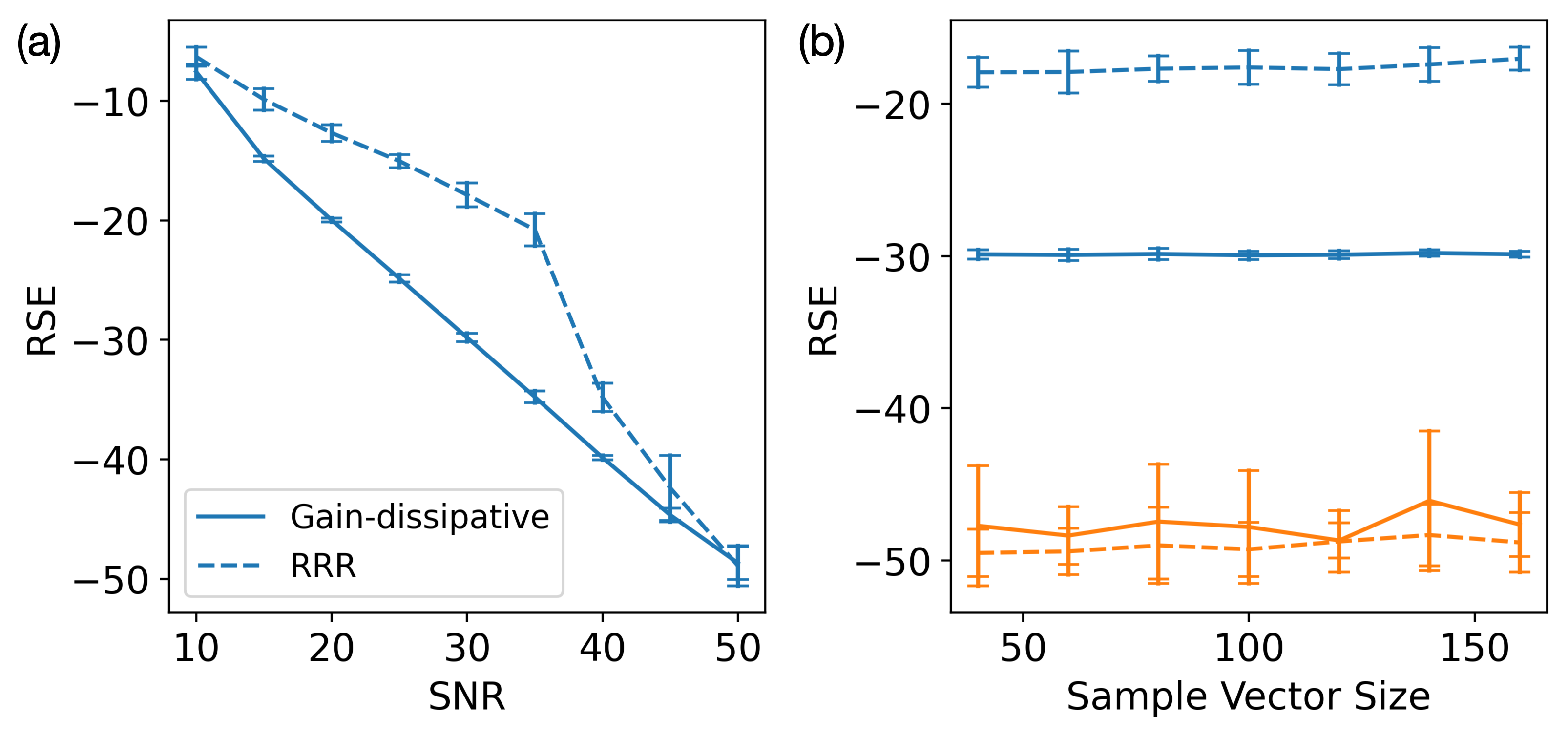}
    \caption{ Phase retrieval from random complex-valued samples.
    (a) Phase retrieval error (RSE) produced by the RRR method (dashed lines) and the gain-based system (solid lines) as a function of noise in the measured amplitudes. At each noise level, 20 random complex samples were generated, each with 100 elements whose amplitudes are uniformly distributed in \([0,1)\) and phases in \([0,2\pi)\). The resulting observation vectors were then used for both methods. Vertical error bars indicate the standard deviation in RSE across the 20 trials.
    (b) Phase retrieval error (RSE) versus the dimensionality of the sample vectors, comparing medium noise (\(\text{SNR}=30\), shown in blue) and low noise (\(\text{SNR}=50\), shown in orange). Solid lines again correspond to the gain-based system, and dashed lines correspond to RRR. Each data point represents the average RSE over 20 distinct random instances of the specified dimension, with error bars indicating the standard deviation.
    }   
    \label{fig:4_dense_rrr_sl_compare}
\end{figure}

While the two-dimensional vortex example provides instructive insight into solver behavior at various noise levels, it remains a highly structured sample. To assess robustness on less structured data, we next consider \emph{unstructured} samples \(\mathbf{x}\), whose amplitudes are drawn uniformly from \([0,1)\) and phases from \([0,2\pi)\). Keeping the dimensionality fixed and varying the noise level, we compared the RRR and gain-based methods, as summarized in Fig.~\ref{fig:4_dense_rrr_sl_compare}(a). The results largely mirror the vortex case: in the medium-noise regime, the gain-based solver attains markedly lower RSE than RRR, whereas both methods perform comparably under very high (\(\text{SNR} < 10\)) or very low (\(\text{SNR} > 40\)) noise. These findings indicate that the gain-based approach handles both structured and unstructured samples effectively, consistently outperforming RRR in the medium-noise band. Moreover, its RSE increases in tandem with SNR, suggesting stable performance across different noise levels.

All of the above tests involved samples of size 100. However, realistic applications typically require recovering much larger vectors. To explore how solution accuracy scales with dimensionality, Fig.~\ref{fig:4_dense_rrr_sl_compare}(b) plots RSE against problem size for medium-noise (\(\text{SNR}=30\), in blue) and low-noise (\(\text{SNR}=50\), in orange) conditions, comparing the gain-based method (solid lines) and RRR (dashed lines). In both noise regimes, the RSE remains nearly constant even as the number of sample elements grows by a factor of four. This suggests that noise level influences the gain-based solver’s performance more than the underlying problem dimension.

\begin{figure}
    \centering
    \includegraphics[width=\columnwidth]{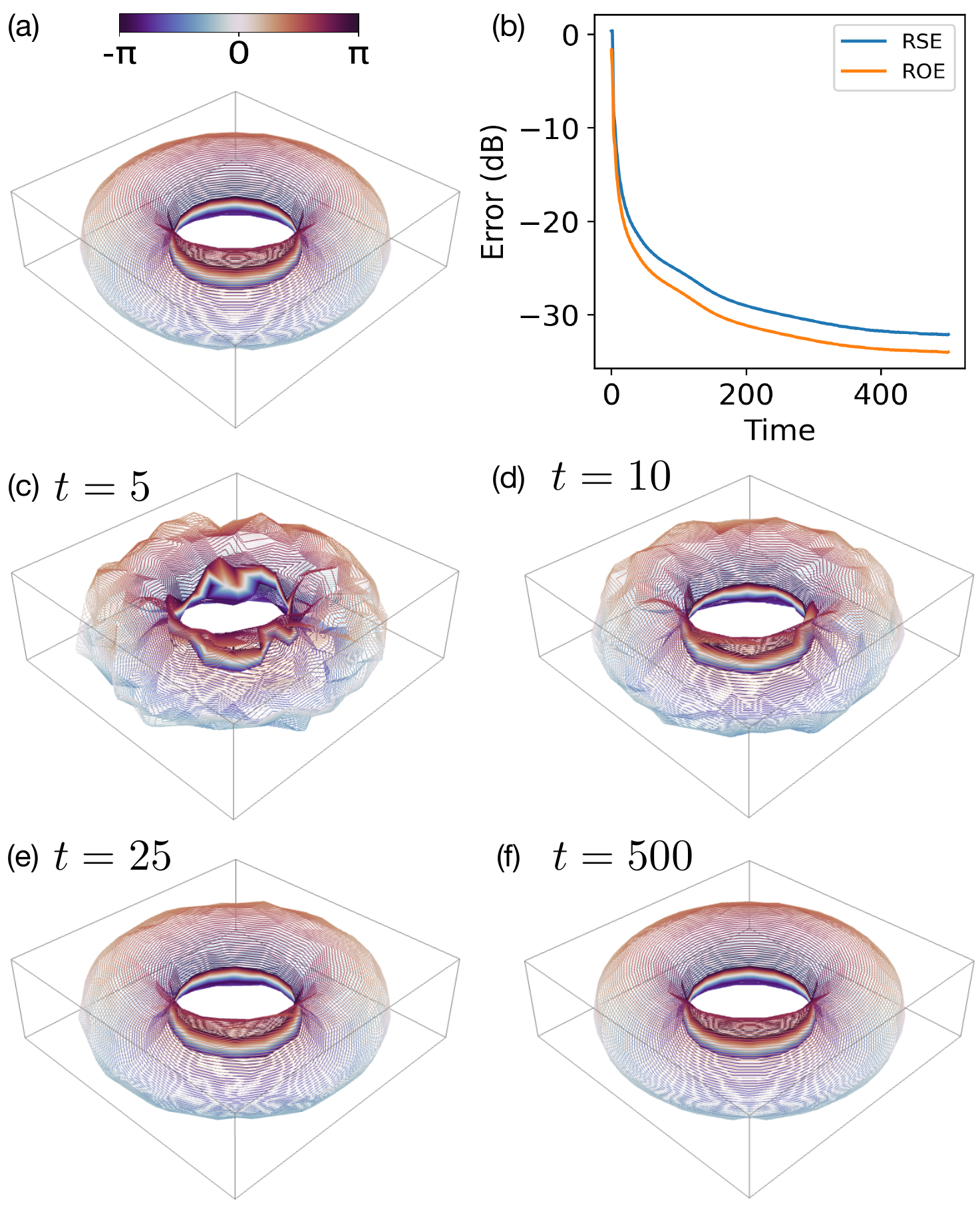}
    \caption{
    Three-dimensional vortex ring reconstructed using the gain-based phase retrieval method.
    (a) Original vortex ring, visualized as an isosurface at 30\% of its maximum amplitude. Phase isolines on this surface are colored according to their phase values, ranging from \(-\pi\) to \(\pi\).
    (b) Time evolution of the phase retrieval error (RSE and ROE) under gain-based dynamics.
    (c)--(f) Snapshots of the reconstructed vortex ring at \(t=5\), \(t=10\), \(t=25\), and \(t=500\), respectively, illustrating the progressive refinement of amplitude and phase in three dimensions.
    }
    \label{fig:7_vortex_ring}
\end{figure}

To illustrate how the gain-based method can recover phase information in cold-atomic Bose--Einstein condensate (BEC) experiments, where topological defects like solitons~\cite{Yefsah2013}, vortex lines~\cite{Serafini_PRX2017}, and vortex rings~\cite{Mark_PRL2014,Bulgac_PRL2014} naturally appear, we consider reconstructing a three-dimensional complex vortex ring. The sample vector is specified by
\begin{equation}
    v(r,\theta,z)
    \;=\;
    \frac{(r-r_0) + i\,z}
         {\sqrt{(r-r_0)^2 \;+\; z^2 \;+\;\xi^2}},
\end{equation}
where \((r,\theta,z)\) are cylindrical coordinates, \(r_0\) is the ring’s radius, and \(\xi\) is the vortex-core size. Figure~\ref{fig:7_vortex_ring}(a) shows an isosurface at 94\% of the maximum amplitude, with phase isolines superimposed. The vortex flow, which is orthogonal to these lines, winds along the ring.

We construct a CDP phase retrieval problem from this 3D complex-valued vector (3087 elements) and solve it using the gain-based system. Figure~\ref{fig:7_vortex_ring}(b) charts the time evolution of the retrieval error, while panels (c)--(f) depict snapshots of the reconstructed three-dimensional field at various stages. By the final state, the key vortex-ring features closely match the original, indicating that, with non-destructive CDP measurements, the gain-based approach can accurately recover the phase of 3D wavefunctions in BEC systems.

\section{Conclusion}
\label{sec:conclusion}

We have shown that the Coded Diffraction Pattern  variant of phase retrieval can be rigorously reformulated as an \textit{XY} Hamiltonian minimization problem, paving the way for direct solution by gain-based oscillator networks and related physics-inspired systems. Through numerical tests, we demonstrated that such gain-based dynamics significantly outperforms the state-of-the-art Relaxed-Reflect-Reflect  algorithm, particularly under medium-level noise (SNR values between 10 and 40\,dB). Our findings hold for both structured data (e.g., two-dimensional vortices and three-dimensional vortex rings) and unstructured complex-valued data with random amplitudes and phases.

Critically, we observed that the superior accuracy of the gain-based solver remains robust even as problem sizes grow. This scalability, combined with its noise resilience, indicates strong potential for large-scale real-world imaging tasks. Moreover, the gain-based approach can be physically realized in optical, polaritonic, or other nonlinear oscillator networks, thereby offering a hardware platform for rapid, energy-efficient phase retrieval. Such physical devices could perform continuous, parallel searches for global minima in the \textit{XY} energy landscape, transforming this theoretical advantage into practical gains for real-time imaging and beyond.

Our results open a promising direction in the development of continuous-variable ``\textit{XY} machines,'' enabling them to tackle large and noisy phase retrieval instances that arise in a variety of scientific and industrial settings.

\section*{Acknowledgements}
The authors acknowledge the support from HORIZON EIC-2022-PATHFINDERCHALLENGES-01 HEISINGBERG Project 101114978.
N.G.B.\ also acknowledges support from Weizmann-UK Make Connection Grant 142568. 

\bibliography{references}
\end{document}